# Kerr beam self-cleaning on the LP$_{11}$ mode in graded-index multimode fiber

E. Deliancourt[1], M. Fabert[1], A. Tonello[1], K. Krupa[1,3], A. Desfarges-Berthelemot[1], V. Kermene[1], G. Millot[2], A. Barthélémy[1], S. Wabnitz[3,4], and V. Couderc[1,*]

[1] *Université de Limoges, XLIM, UMR CNRS 7252, 123 Avenue A. Thomas, 87060 Limoges, France*
[2] *Université de Bourgogne Franche-Comté, ICB, UMR CNRS 6303, 9 Av. A. Savary, 21078 Dijon, France*
[3] *Dipartimento di Ingegneria dell'Informazione, Università di Brescia, via Branze 38, 25123 Brescia, Italy*
[4] *Istituto Nazionale di Ottica del Consiglio Nazionale delle Ricerche (INO-CNR), via Branze 45, 25123 Brescia, Italy*

*Corresponding author: vincent.couderc@xlim.fr

**We report the experimental observation of Kerr beam self-cleaning in a graded-index multimode fiber, leading to output beam profiles different from a bell shape, close to the LP$_{01}$ mode. For specific coupling conditions, nonlinear coupling among the guided modes can reshape the output speckle pattern generated by a pulsed beam into the low order LP$_{11}$ mode. This was observed in a few meters long multimode fiber with 750 ps pulses at 1064 nm in the normal dispersion regime. The power threshold for LP$_{11}$ mode self-cleaning was about three times larger than that required for Kerr nonlinear self-cleaning into the LP$_{01}$ mode.**

Multimode optical fibers (MMFs) are currently extensively revisited for communication applications, and because they provide a convenient experimental platform for the investigation of complex space-time nonlinear dynamics. Various nonlinear propagation phenomena have been theoretically predicted to occur in multimode fibers since the eighties [1-5], but it is not until recently that some of them were actually observed. Consider, for instance, multimode optical solitons [6] and geometric parametric instability (GPI) [7]. On the other hand, experiments on nonlinear propagation in multimode fibers have recently revealed an unexpected effect that was named Kerr beam self-cleaning. It consists in the reshaping, at high powers, of the speckled output intensity pattern into a bell-shaped beam close to the fundamental LP$_{01}$ mode of a graded index (GRIN) MMF. Such nonlinear beam evolution was observed at power levels below the threshold for frequency conversion, as well as below the self-focusing threshold, with sub-nanosecond to femtosecond pulses propagating in the normal dispersion regime [7-10]. It is generally admitted today that Kerr beam self-cleaning results from a complex nonlinear coupling, or four-wave mixing (FWM) interaction among a large population of guided modes. Namely, the combination of spatial self-induced periodic imaging and Kerr nonlinearity creates a periodic longitudinal modulation of the refractive index of the fiber core. This permits quasi-phase matching and energy exchange between guided modes by means of FWM [11]. The nonlinear energy exchange between the fundamental mode and the high-order modes (HOMs) exhibits a nonreciprocal behavior, driven by self-phase modulation [8,12]. In these conditions, all the energy transferred in the fundamental mode remains definitively trapped, which explains the robust nature of the self-cleaning process. Besides that model, alternative concepts based on instability of the HOMs [10], or on modified wave turbulence theory [4] have been introduced, which could also explain the unconventional spatial dynamics observed in MMFs.

In this Letter, we report the experimental demonstration of Kerr beam self-cleaning in favor of the LP$_{11}$ mode of a gradient index (GRIN) MMF. This is quite unexpected, since all previous works on Kerr beam self-cleaning, based on either numerical investigations or experiments, reported self-cleaning of the output field into a smooth, bell-shaped beam, close to the profile of the LP$_{01}$ fundamental mode of the waveguide [7-10].

The experimental set-up to observe Kerr beam self-cleaning into the LP$_{11}$ mode is quite straightforward. It is based on an amplified microchip Q-switched Nd:YAG laser, delivering 750 ps pulses at 1064 nm (quasi single frequency) at a 27 kHz repetition rate, with up to 1 Watt average power. In order to adjust the input power, the laser beam passed through a half-wave plate and a polarizing cube before being focused within the central fiber axis onto the MMF input face. The beam spot was of Gaussian shape and ∼30 μm in diameter full width at half maximum in intensity (FWHMI). The GRIN MMF of 8.3 m in length was loosely coiled on the table forming rings of ∼15 cm diameter. The fiber had a circular core of 26 μm radius, and the core-cladding index difference corresponded to a numerical aperture (NA) of 0.2. The refractive index profile was measured to be smooth, and it could be fitted well by a quadratic law. The fiber carries up to 110 modes per polarization component at 1064 nm. The set-up included three axes precision translation stages for the fiber holders, and for the characterization of laser light delivered by the MMF, a power meter, a polarizer, near-field and far-field monitoring cameras, and a spectrum analyzer. In our experiments, we systematically varied the combination of guided modes excited at the input, in order to verify whether the Kerr beam self-cleaning was always leading to a central bell-shaped beam (which can be associated to a strong enhancement of the energy carried by the $LP_{01}$ mode) at the output of a fixed length of MMF.

In a first series of experiments, we excited the MMF at normal incidence on the fiber axis, taking care to avoid any tilt angle or lateral shift of the input pump beam with respect to the fiber axis or central position (see Fig.1(a)). In that configuration, in principle even parity modes only are excited (with a predominant energy fraction into the $LP_{01}$ mode), owing to the cylindrical symmetry of the input Gaussian beam and the perfect on-axis excitation. This hypothesis was later confirmed by comparing the output beam pattern obtained both in the experiment and in full numerical simulations obtained after only 1 cm of propagation in the MMF (see Fig.1(b)). The numerically calculated modal decomposition of the input beam is also presented in Fig.1(c).

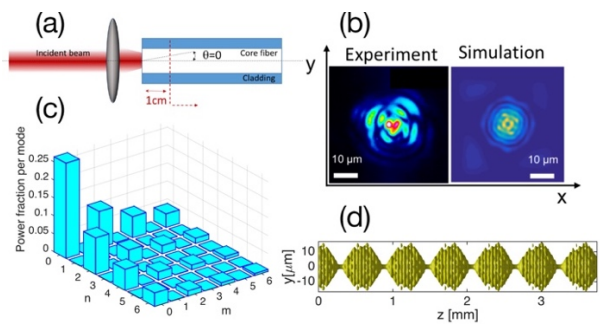

Fig. 1: (a) Experimental condition used for axial power coupling in the GRIN-MMF fiber, (b) Corresponding experimental and numerical near field intensity patterns after 1cm of propagation in the GRIN-MMF, (c) Fraction of power coupled into the guided modes (Hermite-Gauss basis); (d) Iso-intensity surface (at 50% of the local maximum intensity) upon the propagation distance z.

The spatial coherent beating between excited even modes creates a periodic local intensity oscillation along the fiber, whose transverse intensity is symmetrically distributed around the fiber axis, exhibiting on-axis peak intensity localization. Such symmetrical intensity oscillation in combination with the Kerr effect induces a modulation of the local refractive index with a similar shape (see Fig.1(d)), which in turn permits the quasi phase-matching of nonlinear mode coupling, mainly for those modes with peak intensity in the core center. The cleaning process appears to favor the fundamental mode, which has, in this case, the smallest effective area. Thus, when operating with an on-axis initial conditions, we reproduced the previously reported Kerr self-cleaning into a quasi-Gaussian (bell-shape) distribution at the MMF output (at 8.3 m), with an efficient energy transfer towards the fundamental mode [8] (see Fig. 2(a)).

In a second series of experiments, we varied the input tilt angle of the Gaussian laser beam (see Fig. 3(a)). The incident external angle was chosen to be close to 2.5°, in order to excite the fiber beyond the numerical aperture of the fundamental mode. In doing so, we limited the amount of energy coupled into that mode. This configuration allowed to excite a combination of even and odd modes, now with the highest fraction of power coupled into the $LP_{11}$ mode (see Fig. 3(c)). The comparison of the measured and calculated output beam patterns recorded after 1 cm of propagation in the GRIN MMF provides an indirect information about the initial input conditions achieved in practice, and more specifically on the associated modal content. Figure 3(c) shows the numerically calculated power fraction coupled in each transverse mode, which corresponds to the simulated output beam shape illustrated in panel (b) of the same figure. This modal distribution generates a Kerr-induced longitudinal refractive index grating, whose peak intensity positions exhibit an off-axis transverse localization in the form of a *zig-zag* trajectory around the fiber axis, in the plane defined by the angle of the incident beam (Fig. 3(d)). Such a spatial geometry of the refractive index modulation is expected to lead to a strong overlap with the $LP_{11}$ mode which, in this second configuration, should favor quasi phase-matching of FWM processes involving that mode. The resulting experimental output beam shape at 8.3 m is shown in Fig. 2(b).

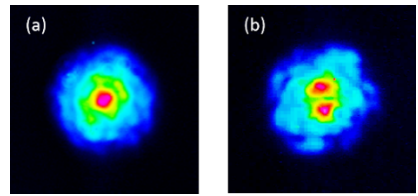

Fig. 2. Intensity patterns at the GRIN-MMF output recorded (a) for standard Kerr self-cleaning on a bell-shaped beam at 2kW peak power, and (b) for Kerr self-cleaning on a $LP_{11}$ profile, for specific beam coupling conditions and 4.5kW peak power.

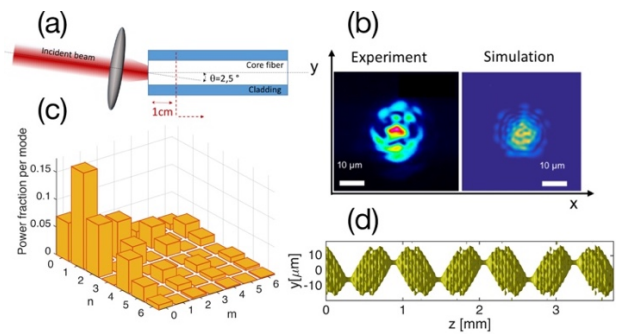

Fig. 3. (a) Experimental condition used for power coupling in the GRIN-MMF fiber with an input incidence angle of 2.5°; (b) Corresponding experimental and numerical near field intensity patterns after 1cm of propagation in the GRIN-MMF, (c) Fraction of power coupled into the guided modes (Hermite-Gauss basis), (d) Iso-intensity surface at 50% of the local maximum intensity along the propagation distance z.

To obtain the experimental confirmation of spatial self-cleaning on the $LP_{11}$ mode, we proceeded as follows: we started with high peak power laser pulses, and we adjusted the tilt angle of the input beam with respect to the fiber axis, until we observed a nonlinear self-cleaning of the LP11 mode. This self-organization was obtained at the expenses of a moderate reduction of the coupling efficiency with respect to that involving the $LP_{01}$ mode (see Fig. 3).

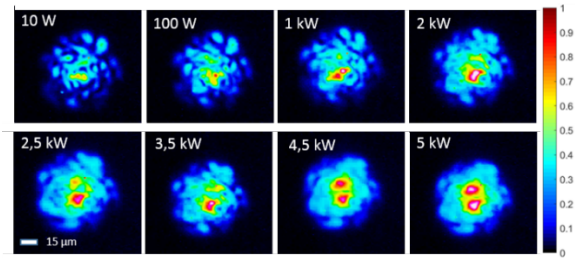

Fig. 4. Near field intensity patterns at the GRIN-MMF output recorded for increasing peak power values at the input. Laser beam coupling into the MMF was appropriate for Kerr self-cleaning on a $LP_{11}$ profile.

The input laser power was then progressively decreased, until the linear propagation regime was recovered, and a highly-speckled beam profile was observed (see visualization 1). In that way, it was possible to confirm that nonlinear selection of the $LP_{11}$ mode at the fiber output did not simply result from the excitation of that unique mode from the very beginning of propagation in the MMF. The recorded fiber output images, for a given input coupling condition, are presented in Fig.4, for increasing input beam power values. The threshold peak power to obtain Kerr self-cleaning in the LP11 mode is of the order of 4 - 5 kW. This is a value close to, but larger than the ~1kW power which is necessary to observe, in the same experimental conditions, self-cleaning into the $LP_{01}$ mode. By supposing that the observed double-peaked output intensity pattern, obtained as a result of nonlinear propagation in the GRIN MMF, indeed corresponds to the $LP_{11}$ mode, one would conclude that the field profile is described by a Laguerre Gauss (LG) distribution, with two lobes of opposite phases. Because of the mathematical properties of the Fourier transform of LG functions, the far field associated with the $LP_{11}$ mode is also a LG distribution of the same type. Therefore, it should exhibit an intensity profile which is similar in shape to that of the near field. This is indeed what we observed, with a simultaneous display on two cameras placed at the fiber output in the near field and in the far field, respectively.

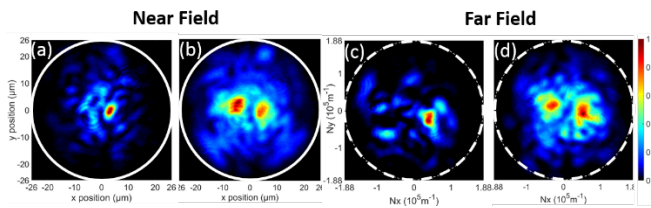

Fig. 5. Near field (a),(b) and far field (c),(d) intensity patterns at the GRIN-MMF output recorded in linear propagation regime (a, c) and in the nonlinear self-cleaning regime (b, d) for appropriate settings of the input coupling, in order to get self-cleaning of a $LP_{11}$ mode. The white line corresponds to the core boundary in the near field images and to the NA in the far field images.

For example, Fig.5 compares the output near field and far field recorded at low powers (*i.e.*, in the purely linear propagation regime) with those recorded at high input powers, leading to a self-cleaning behavior into the $LP_{01}$ mode. From Fig.5, it is clear that nonlinear propagation in the MMF altered the guided beam mode expansion, leading to an apparent reshaping of both the near field and far field at the fiber output. We may notice in Fig.5 the two-lobe structure of both the far and the near fields, with a similar angular orientation. This supports the claim that self-cleaning of the $LP_{11}$ mode has indeed been achieved. In order to provide a more quantitative picture of the nonlinear self-cleaning on the $LP_{11}$ mode, one may go beyond the simple observation of the fields delivered by the GRIN-MMF. We processed the image recordings to compute an intensity correlation parameter, $C_s$. This parameter is defined as the integration on the cross-section (dS stands for the surface element) of the normalized product of the surface of the recorded ($I_{exp}$) and theoretical ($I_{th}$) $LP_{11}$ pattern, determined by their iso-lines at half maximum in intensity:

$$C_S = \frac{\int I_{exp} I_{th} dS}{\sqrt{\int I_{exp}^2 dS \int I_{th}^2 dS}} \qquad (1)$$

The evolution of $C_s$ versus launched power is given in Fig.6 for both the near field and the far field (or plane wave spectrum). The two curves in Fig.6 indicate that the correlation between the experimentally observed patterns and the $LP_{11}$ mode steadily grows when nonlinear mode coupling is stronger, *i.e.,* when the laser power is raised. This evolution testifies that the observed modal self-cleaning can be significantly stable for a wide range of input pump powers. Although the parameter $C_s$ was computed from intensity measurements, the phase information of the near field is encoded in the intensity pattern of the far field. This is the reason why we provided the correlation parameter for both near and far fields. The similar evolution of the two correlations means that the phase as well as the intensity of the observed output progressively approaches that of a $LP_{11}$ mode, as the input power was increased. Hence the power fraction into the $LP_{11}$ was enhanced at high power. We verified that Kerr self-cleaning into the $LP_{11}$ mode occurred at power levels such that self-phase modulation did not significantly broaden the laser pulse spectrum.

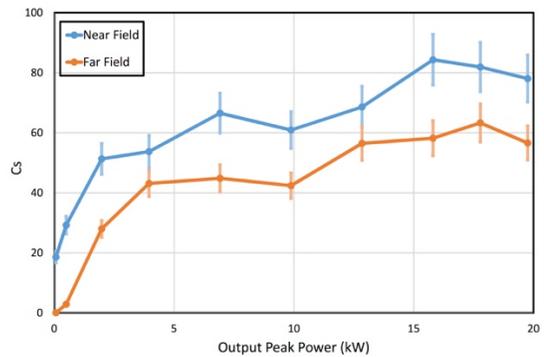

Fig. 6. Correlation between the experimental patterns at the GRIN-MMF output and the $LP_{11}$ theoretical shape, upon the launched laser power, in the near-field (blue curve) and far-field (red curve), respectively.

Moreover, at these power levels frequency conversion via either Raman scattering, intermodal four-wave mixing, or GPI was not observed. Nonlinear self-cleaning in favor of the $LP_{11}$ mode is robust with respect to external perturbations, similarly to what already previously reported for Kerr beam self-cleaning of the fundamental or $LP_{01}$ mode [8]. When the fiber loops were shaken, bent or squeezed by hands, we observed fluctuations in the output image, especially in the background speckled patterns, but the main two lobes structure remained well preserved in shape and orientation (see visualization 2).

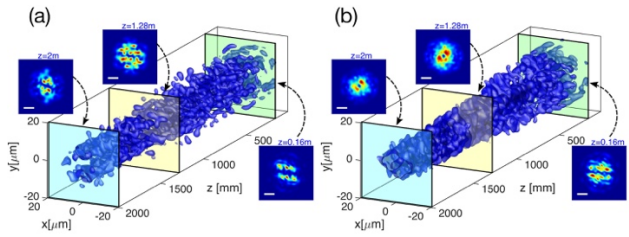

Fig. 7. Time-averaged numerical results (Iso-intensity surfaces at 50% of the local maximum) of beam propagation in a GRIN MMF. (a) absence of Kerr effect. (b) Kerr effect enabled. The insets show the beam intensities at three different positions along the propagation. White segment: 10 µm.

To reproduce our experiments, we carried our numerical simulations by solving a 3D vector nonlinear Schrödinger equation, using a coarse step every 5 mm to mimic mode coupling [8]. The simulated fiber core shape was made slightly elliptical (with an incertitude of ±0.1 µm for both transverse axes) and randomly oriented. The electric field was also randomly rotated, and a fixed linear birefringence of $5\times10^{-7}$ was applied on each segment. In doing so, speckles were gradually formed along the course of the numerical propagation, even if the input condition was a Gaussian beam with a diameter of 40 µm, with an additional spatial phase shift of π for y>0, and a pulse duration of 5 ps. The peak intensity was 5GW/cm$^2$. The numerical results reproduce fairly well our experimental observations, showing a re-shaping into the $LP_{11}$ mode (see Fig.7). The absence of input phase shift led instead to cleaning on a bell-shaped mode close to $LP_{01}$ (not shown). In our simulations, we did not include the Raman effect, and the spectrum was of only 30 nm.

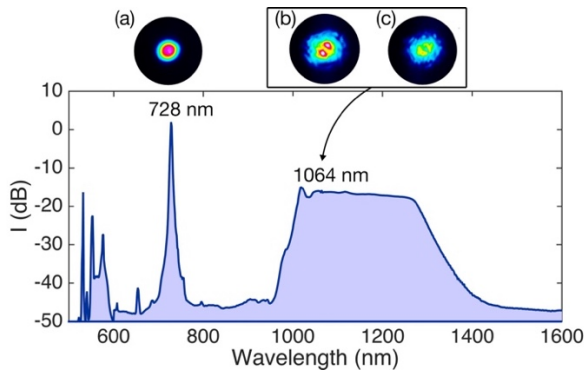

Fig. 8. Experimental output spectrum for the maximum input peak power (50 kW) coupled in the GRIN MMF within a 2.5° input angle. (a) output beam profile of the main intermodal FWM process at 728 nm, (b) output self-cleaned beam at 1064 nm before Raman and intermodal FWM generation (38 kW), (c) output pump beam pattern at 1064 nm after Raman and intermodal FWM generation; fiber length: 6 m.

In order to experimentally analyze the ultimate stability of Kerr self-cleaning into the $LP_{11}$ mode upon large variations of input power, we repeated the experiment with the same input beam tilt (θ= 2.5°), but with significantly higher input pump powers up to 50 kW. Again, we obtained Kerr self-cleaning on the $LP_{11}$ mode, but we were unable to detect any signature of possible decay of the $LP_{11}$ mode, for example towards the fundamental mode. For the maximum peak power that we could launch into the fiber (~ 50 kW), only a blurring of the output beam shape was observed.

The spectral analysis revealed that additional nonlinear processes, *i.e.* intermodal FWM and Raman effect, significantly depleted the cleaned beam. Evidence of the nonlinear frequency conversion processes, effectively limiting the maximum power carried by the input pump beam, is shown in Fig. 8. Note that the intermodal FWM signal sideband is obtained at 728 nm (the idler sideband is expected at 1976 nm) on the fundamental transverse mode, whereas the pump wave at 1064 nm still emerges in the $LP_{11}$ mode.

To conclude, we have experimentally shown that Kerr beam self-cleaning in GRIN-MMFs can reshape the transverse output pattern into the $LP_{11}$ mode of a GRIN MMF, starting from a broad speckled pattern in the linear propagation regime. Nonlinear self-cleaning of the $LP_{11}$ mode requires a careful adjustment of the laser beam coupling at the fiber input, in order to prepare a proper power distribution among the guided modes. Numerical simulations are consistent with our experimental results, and the beam shape is robust against fiber squeezing and bending. As the input power grows larger, power depletion caused by parametric sideband generation and Raman scattering eventually limits further power increases of the output pump beam. Our observations should stimulate further research on spatio-temporal self-organization processes in MMFs. Moreover, the possibility of engineering a family of robust nonlinear spatial attractors from MMFs may have important applications in the delivery of high-power laser beams for micro-machining and nonlinear microscopy applications.

**Funding.** The European Research Council (ERC) under the European Union's Horizon 2020 research and innovation programme (No. 740355). K.K. has received funding from the European Union's Horizon 2020 research and innovation programme under the Marie-Skłodowska-Curie (No. 713694), The french "Investissements d'Avenir" program, project ISITE-BFC (contract ANR-15-IDEX-0003)